# Extracellular biosynthesis of silver nanoparticles from Plant Growth Promoting Rhizobacteria *Pseudomonas sp.*

Archana Yadav[1], T. Theivasanthi[2], P.K. Paul[1] and K.C. Upadhyay[1]

[1]Amity Institute of Biotechnology, Amity University, Noida, India.
[2]International Research Center, Kalasalingam University, Krishnakoil, Tamilnadu.
Emails: ttheivasanthi@gmail.com, ayadav86@amity.edu

___________________________________________________________________________

**Abstract:** Development of reliable and eco-friendly processes for synthesis of metallic nanoparticles is an important step in the field of application of nanotechnology. In the present study, extracellular synthesis of silver nanoparticles (AgNPs) using culture supernatant of plant growth promoting rhizobacterium, *Pseudomonas sp.*ARS-22. The biosynthesis of AgNPs by *Pseudomonas sp.*ARS-22(denoted as arsAgNPs) was monitored by UV–visible spectrum that showed the surface plasmon resonance (SPR) peak at 410 nm, an important characteristic of AgNPs. Furthermore, the morphological, elemental, functional and thermal characterization of arsAgNPs was carried out using the electron and atomic microscopies, FTIR spectroscopy and Atomic Force microscopy (AFM), respectively. The arsAgNPs were spherical in shape. The Debye–Scherrer equation was used to calculate particle sizes and the silver nanoparticles of approximate size 10-30 nm were observed. The process of reduction is extracellular, which makes it an easier method for the synthesis of silver nanoparticles.

___________________________________________________________________________

## 1. Introduction

Human population is growing day by day and food supply for each is the major problem in the present scenario. To overcome this problem with a scientific gateway; improving the crop yield with the application of modern biotechnological tools and techniques in practice can be undertaken. One of the important modern technologies which are the demand of present scenario also is use of Nanotechnology in the field of agriculture. A different nanoparticles plays their important role in the improvement of crop yield by several mechanisms as enhances plant growth by suppressing the diseases, antibacterial as well as growth promoter and in tissue engineering etc. Nanoparticles are also playing significant role in our daily life as several commercial products are made for human beings. Nanotechnology is the understanding and control of matter at dimensions between approximately 1 and 100 nanometers, or nano-scale. Unusual physical, chemical, and biological properties can emerge in materials at the nano-scale. These properties may differ in important ways from the properties of bulk materials and single atoms or molecules (Bakshi *et al,* 2014). Nanotechnology involves imaging, measuring, modelling, and manipulating matter at this length scale (encompassing nano-scale science, engineering, and technology).

Biological synthesis of nanoparticles has emerged as rapidly developing research area in nanotechnology across the globe with various biological entities being employed in production of nanoparticles constantly forming an impute alternative for conventional methods. In natural biological systems scientists developed an alternative strategy for nanoparticles synthesis using microorganisms (Malhotra *et al,* 2013). Given the high surface area relative to the amount of nano-materials, fertilizers based on nanotechnology have the potential to surpass conventional fertilizers, but lack of funding and concern over regulation and safety has led to little advancement in this area. In nano-fertilizers, nutrients can be encapsulated by nanomaterials, coated with a thin protective film or delivered as emulsions or nanoparticles. Nanomaterials could even be used to control the release of the fertilizer such that the nutrients are only taken up by the plant and not lost to unintended targets like soil, water, or microorganisms.

Microbial synthesis of nanoparticles is eco-friendly and has significant advantages over other processes since it takes place at relatively ambient temperature and pressure. As the size and shape of nanoparticles can also be controlled in microbial synthesis, screening of unexplored microorganisms for AgNPs synthesizing property is very important.

Plant growth promoting *rhizobacteria* are a group of non pathogenic, occurred mainly in *rhizosphere* and associated with root legumes. They can improve crop yield by different mechanisms as they can fix atmospheric nitrogen into its simplest form so that plant can use nitrogen. These bacteria fixed nitrogen by symbiotically and non-symbiotically. *Rhizobium, Pseudomonas, Azospirillum, Azomonas, Xanthobacter and Bacillus* etc. are some well known nitrogen fixing bacteria. When plants are growing naturally in soils, one cannot differentiate whether clear growth promotion is caused by bacterially stirred plant growth or through inhibition of lethal soil microorganisms. Non-pathogenic *rhizobacteria* can antagonize pathogens. Some PGPR takes such type of action against plant pathogen as these bacteria are native of *rhizospheric* soil where pathogenic fungi are attracted to plant roots. However, *rhizobacteria* can reduce the activity of pathogenic microorganisms not only through microbial antagonism but by increasing the self defense mechanism of plant. *Pseudomonas fluorescens* and *Bacillus* sp. are excellent examples of so called biocontrol mechanism. Plant growth promoting *rhizobacteria* are also playing an eminent role during stress conditions. The synthesis and activity of nitrogenases in *A.brasilense* is restricted by salinity stress and ACC deaminase for enhancement of plant under unfavourable environmental conditions such as flooding, heavy metals, phytopathogens, drought and high salt is a prominent case of PGPR. *Pseudomonas* sp. is ever-present bacteria in agricultural soils and has many characters that make them well suited as PGPR. The most effective strains of *Pseudomonas* have been *Fluorescent Pseudomonas* sp. Considerable research is under way globally to develop the potential of one group of bacteria that belongs to *Fluorescent pseudomonads* (FLPs). FLPs help in the maintenance of soil health and are metabolically and functionally most dissimilar.

## 2. Materials and Methods

**Isolation and Identification of Bacterial Strain ARS-22**
Bacteria were isolated from soil samples collected from agricultural fields of the Amity University Campus, NOIDA, INDIA. The soil samples were subjected to serial dilution and plated on Nutrient Agar (NA) medium. After incubation at 30°C for 24 h, bacterial colonies were subcultured and further purified on NA. On the basis of rapid reduction of silver nitrate (AgNO$_3$) into AgNPs, the bacterial strain ARS-22 was selected for further studies. The preliminary characterization of the bacterial strain ARS-22 based on physiological and biochemical characteristics was carried out according to Bergey's Manual of Determinative Bacteriology (Holt *et al*. 1994). The qualitative plant growth promoting activities of the strain ARS-22 were also ascertained. The inorganic phosphate solubilizing ability was determined using NBRIP-BPB medium as described earlier (Mehta and Nautiyal 2001). The indole acetic acid (IAA) production was determined according to the method described by Brick *et al*. (Brick *et al*. 1991) and siderophore production was tested using Chromazural S(CAS) assay as described by Dwivedi *et al*. (Dwivedi *et al*.2011).

**Extracellular Biosynthesis of AgNPs by *Pseudomonas sp*. ARS-22**
The bacterial strain **ARS-22** was inoculated in Nutrient Broth (NB) medium and incubated at 30uC for 48 h to attain the early stationary phase (Mussarat *et al*. 2010). The culture supernatant was obtained by centrifugation at 10,000 rpm for 10 min and transferred to another sterile conical flask. The fresh stock of AgNO$_3$ was prepared in sterile distilled water and added to the culture supernatant at the final concentration of 1 mM. The conical flask was incubated in dark and the synthesis of AgNPs was monitored by visual colour change from yellow (original colour of culture supernatant) to dark brown (colour of culture supernatant after adding AgNO$_3$) (Cason *et al*., 2000; Nair *et al*., 2010). The arsAgNPs were air dried in sterile conditions and obtained in the form of powder for further studies.

**Characterization of arsAgNPs**
The optical, morphological, elemental and functional characterization of the arsAgNPs was carried out using the UV–vis spectrophotometer, electron and atomic microscopes, (EDAX) spectrometer, Fourier transform infrared (FTIR) spectroscopy and atomic force microscopy (AFM), respectively. In order to ascertain the optical characteristics of arsAgNPs in reaction solution, the absorption spectrum was recorded by UV–visible spectrophotometer (Schmidzu) in the wavelength range of A200 to A600 nm using quartz cuvette. The scanning electron microscopy (SEM) was carried out using fine powder of the arsAgNPs on a carbon tape in JSM 6510LV scanning electron microscope (JEOL, Tokyo, Japan) at an accelerating voltage of 20 kV. The elemental analysis of arsAgNPs was done using Oxford Instruments INCAx-sight EDAX spectrometer equipped SEM. The transmission electron microscopy (TEM) of arsAgNPs was carried out on JEOL 100 / 120 kV TEM (JEOL, Tokyo, Japan) with an accelerating voltage of 200 kV. For atomic force microscopy, a thin film of arsAgNPs was prepared on the borosilicate glass slide to analyse the surface morphology. The prepared thin film was analysed on the atomic force microscope (AFM; Innova SPM, Veeco) in the tapping mode. The commercial etched silicon tips were used as scanning probes with typical resonance frequency of 300 Hz (RTESP, Veeco). The microscope was placed on a pneumatic anti-vibration desk, under a damping cover and the analysis was done using SPM Lab software (Petit *et al*.1993; Raether1977). The images of electron microscopes of EDAX and AFM were obtained and converted into an enhanced meta-file format. For the functional characterization of the arsAgNPs, the powder was mixed with spectroscopic grade potassium bromide (KBr) in the ratio of 1:100 and spectra was recorded in the wave number range 400–4000 cm$^{-1}$ on PerkinElmer FT-IR spectrometer Spectrum (Perkin Elmer Life and Analytical Sciences, CT, USA).

## 3. Results and Discussions

**Electron microscopic studies**
SEM analysis was carried out to understand the particle shape of the bio-synthesised silver nanoparticles. It shows that the prepared particles are in spherical shape (Fig.1.a.). Some of the small Ag nanoparticles have been indicated. TEM analysis results are in (Fig.1.b.) corroborate with SEM result *i.e.* the particles are in spherical shape, quiet monodispersed and evenly separated. The results also confirm that the well dispersed particles are in nano range with smooth surface. The particles diameter is in the size range of 10-30 nm. The clear and uniform lattice fringes in HRTEM image confirm highly crystalline nature of the sample. The HRTEM image of single silver nanoparticle of the sample is presented in (Fig.2.a.). It gives us further insight into the structure and crystallinity of the prepared silver nanoparticles. It clearly shows the FCC (Face Centered Cubic) structure. The lattice spacing of 0.25 nm corresponds to (111) planes of silver. The results show that (111) is the dominant face of silver spheres. The ring-like structure in selected area electron diffraction (SAED) pattern of TEM indicates that the particles are crystalline nature (Mani *et al*. 2013). SAED pattern of the sample silver NPs is presented (in Fig.2.b.). It shows the ring pattern which confirms the crystalline nature of the sample. The diffraction rings have been indexed on the basis of the FCC structure of silver. These rings are due to the reflections from (111), (200), (220), (311) and (222) lattice planes of silver. Both HRTEM image and SAED pattern confirmed that the prepared spherical silver nanoparticles are in crystalline nature.

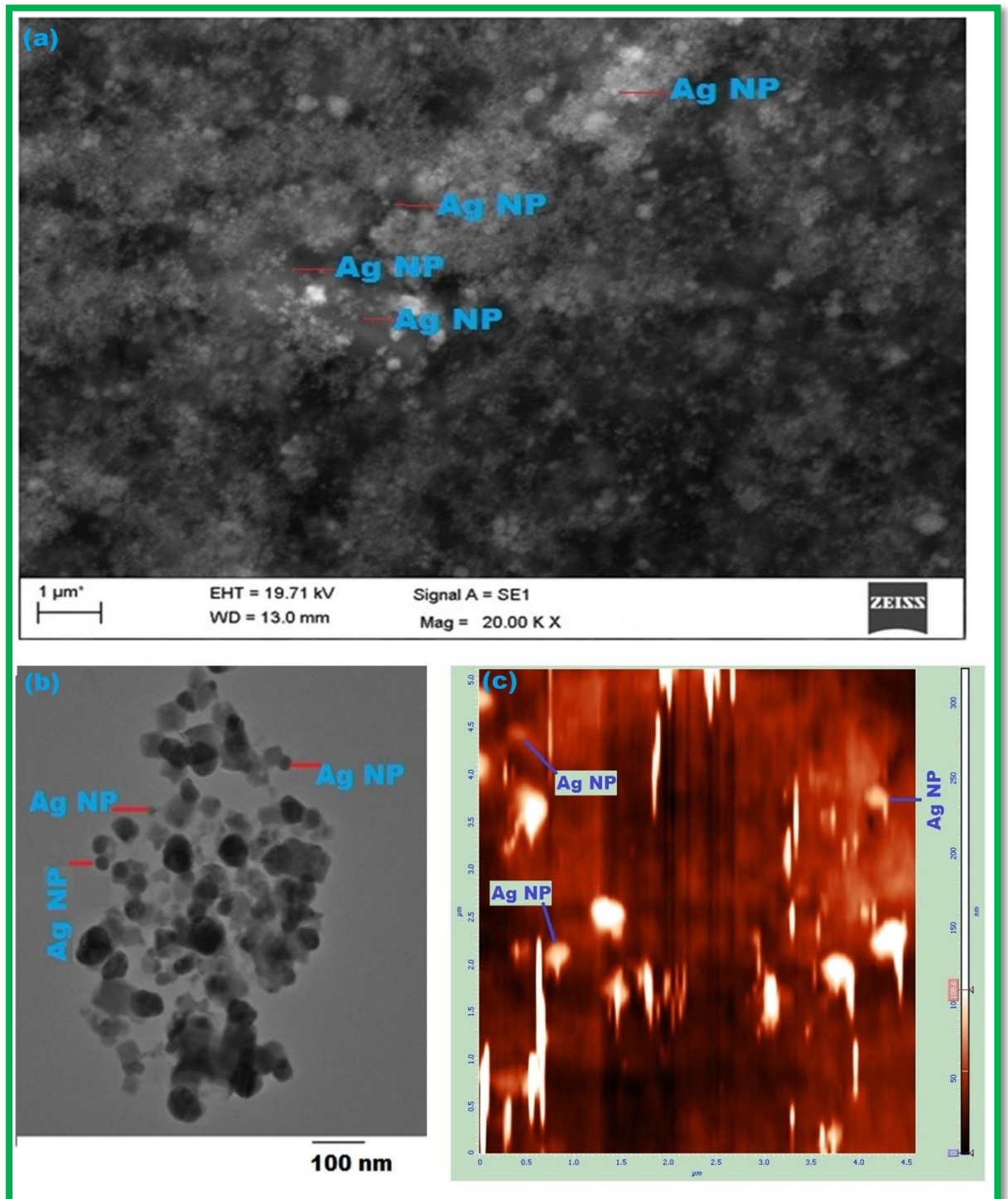

**Fig.1** Bio-synthesized silver nanoparticles. Images of (a) SEM. (b) Bright field TEM. (c) AFM.

**Atomic force microscopic studies**
Surface topology of the prepared sample was studied by atomic force microscopy (AFM) analysis. It is observed from the results that the formulated particles possess spherical shape. Some particles are indicated for easier identification. AFM picture of the sample has been shown in fig. 1.c. It shows that the particles size is bigger and the size value is greater than TEM measurements. Alahmad *et al*. reports some reasons for such appearance *i.e*. the silver nanoparticles tend to form aggregate / cluster together on the surface during deposition for AFM analysis. Also, the shape of the tip AFM may cause misleading cross sectional views of the sample. So, the width of the nanoparticle depends on probe shape (Alahmad *et al*. 2013).

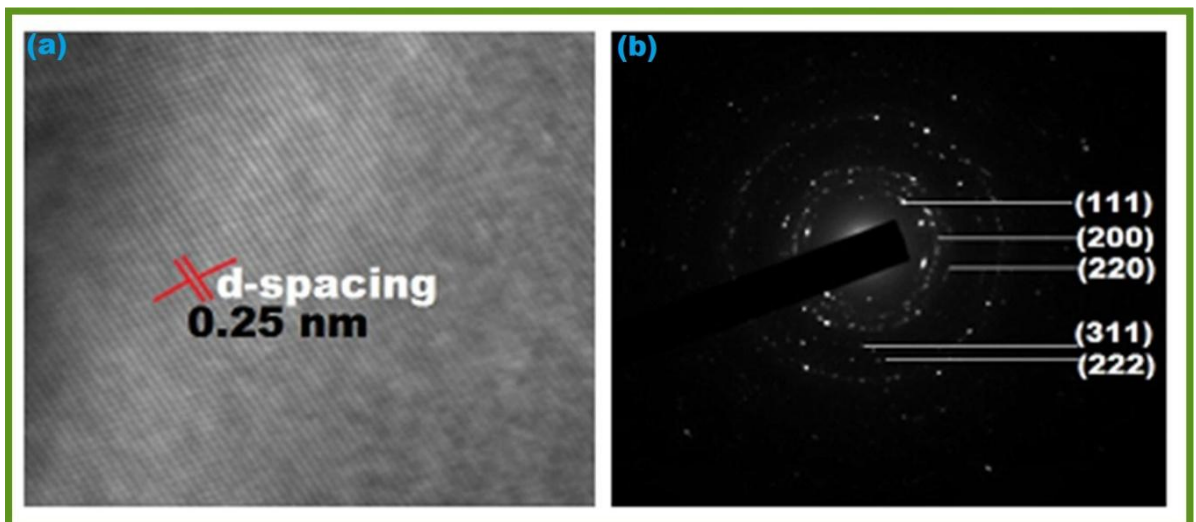

**Fig.2** HRTEM images of bio-synthesized silver nanoparticle. (a) lattice fringes (d-spacing). (b) SAED Ring pattern of FCC

**FT-IR spectroscopic studies**

FT-IR spectroscopic studies were carried out to investigate the functional groups and Ag nanoparticles formation mechanism particularly to identify possible interaction between silver precursor salt and protein molecules, which leads the reduction of silver ions and stabilization of silver nanoparticles (Theivasanthi and Alagar 2013; Poerter *et al*. 1987). The representative spectrum of silver nanoparticles is shown in Fig 3. It exhibits absorption peaks at wavelengths 459, 1336, 1388, 1409, 1457, 1506, 1588, 1636, 1653, 1670, 2362 and 3444 cm$^{-1}$.

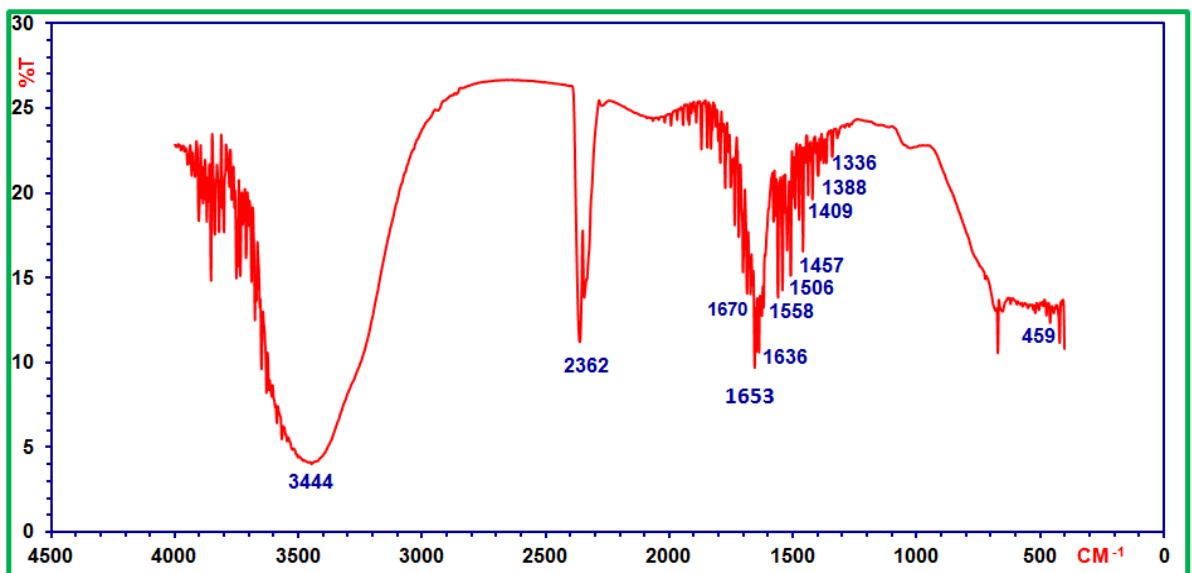

**Fig.3** FT-IR spectrum of silver nanoparticles

The absorption peaks around 1600 and 1350 cm$^{-1}$ represents the presence of $NO_2$ which may be from $AgNO_3$ Solution, the metal precursor involved in the Ag nanoparticles synthesis process. Strong interaction of water with the surface of Silver could be the reason for the O-H group which is represented by peaks around 1350 cm$_{-1}$ (Nakamoto 2006; Goswami and Sen 2004). The peak at 3441 cm$^{-1}$ contains OH stretching modes.

The carboxyl / carbonyl groups and $NO_2$ groups (such as nitro compounds, nitrates and nitramines) commonly exhibit vibrations at 1660 to 1500 and 1390 to 1260 $cm^{-1}$ region (Khan *et al*. 2011; Augustine and Rajarathinam2012). The peak at 1558 $cm^{-1}$ indicates the formation of metal carbonyl group due to the stabilization of Ag nanoparticles by –coo-group (Giri *et al*., 2010) .The protein might play an important role in the stabilization of silver nanoparticles. The presence of bands at 1653 and 1637 $cm^{-1}$ are due to carbonyl stretch in the amide linkages of the proteins.

The FTIR result thus indicates that the secondary structure of proteins is not affected as a consequence of reaction with the $Ag_+$ ions or binding with Ag nanoparticles (Logeswari *et al*. 2013; Jeevan *et al*. 2012). It suggests that the biological molecules could possibly perform a function for the formation and stabilization of Ag NP in an aqueous medium. Therefore, stabilization of the Ag NP by surface-bound proteins is possible one.

It is well known that proteins can bind to Ag NP through free amine groups of proteins (Theivasanthi *et al*. 2011; 2013). The bands at 1387 and 1457 $cm^{-1}$ indicate C-N stretching vibrations and amine /amino-methyl groups respectively. The band seen at 1639 $cm^{-1}$ is characteristic of –C=O carbonyl groups and –C=C- stretching. The overall observation confirms the presence of protein in samples of silver nanoparticles. It reports that protein can bind to nanoparticles either through their free amine groups or cysteine residue and stabilize them. The wave numbers related to amines, amides and amino acids in FTIR spectrum indicate the presence of protein (Gopinath *et al*., 2013; Ramajo *et al*.2009).

The amino acid residues in polypeptides and proteins give well known signature in the infrared region of the electromagnetic spectrum. The amide 1 and 2 of protein secondary structure bands in the FTIR spectrum are sensitive indicators of conformational changes. The amino group is one of the key factors in controlling the metal nanoparticles (Lin *et al*. 2005). The bands at wave numbers 2362, 1670, 1454, 1400 and 1334 $cm^{-1}$ imply the presence of protein / peptide on the nanoparticle surface. Proteins bind to metal nanoparticles through amine and carboxylates ($COO^-$ at the wave number 1400 and 1334 $cm^{-1}$). The NH/C-O groups also indicate the presence of biomolecules like nitrates and Carboxylic Acids (Das *et al*. 2009; Frattini *et al*. 2005).

The protein material are encapping the metal particles and likely to serve as a capping / stabilizing agent and it is concluded from earlier report NH/C-O groups bind with metals (silver) nanoparticles. Amino acid residues and peptides of proteins have the stronger ability to bind metal and very high affinity to bind with metals. So that protein itself can act as an encapsulating agent and possibly form a layer covering the metal nanoparticles (Joglekar *et al*., 2011). This layer prevents agglomeration, protects the nanoparticles from agglomerization and thereby stabilizes the medium .Some researchers have attributed a key role to the amine in the $Ag^+$ reduction due to its decrease in the potential of $Ag^+/Ag$ ($E_{Ag+}/_{Ag}$) which promoting the reaction. However, there is no substantial evidence that confirms this assumption. Gopinath *et al.* report that the band at 463 $cm^{-1}$ indicates metal (Ag) (Gole *et al*. 2001; Andrews and Ozin 1986).

It is observed from the FTIR spectrum of the prepared sample; the peak 1336 $cm^{-1}$ implies that $AgNO_3$ metal precursor has been utilised in sample preparation; the band at 459 $cm^{-1}$ of sample confirms that the presence of metal (Ag); the bands 1336, 1388, 1506, 1588, 1636 and 1653 $cm^{-1}$ of the sample indicates the formation of metal carbonyl group due to stabilization of Ag nanoparticles and presence of protein; 1409, 1457, 1670 and 2362 $cm^{-1}$ also indicates the protein complex on the sample; the proteins play an important role in controlling and stabilization of sample silver nanoparticles; they are binding with nanoparticles and encapping / encapsulating them; this action forms a layer which protects the nanoparticles from agglomerization (Chander *et al*. 2006).

**UV-Vis Studies**

SPR is created on the boundary of a metal. It represents quantized oscillations of surface charge produced by an external electric field. It excites on the surface of metal nanoparticles and creates interesting optical properties. In metal nanoparticles like gold, silver and copper, occurrence of dipole resonance in the UV-Vis region, making them useful for optical applications (Theivasanthi and Alagar 2013; Dang *et al*. 2011). As well as the dielectric properties of the material, the frequency of the dipole resonance depends upon the size and shape of nanoparticles (Ebbesen *et al*. 1993).

SPR peak is an important and special characteristic feature of metal nanoparticles. Vivekanandhan *et al*. reports that the SPR peak at 430 nm (λmax) confirms the attribution and formation of silver nanoparticles. UV-Vis spectrum (wavelength Vs absorbance) of the prepared sample in Fig.4 shows SPR peak at 425 nm. The presence of SPR confirms metallic nature *i.e.* the prepared sample is Ag metal nanoparticles. It also corresponds to the previous report.

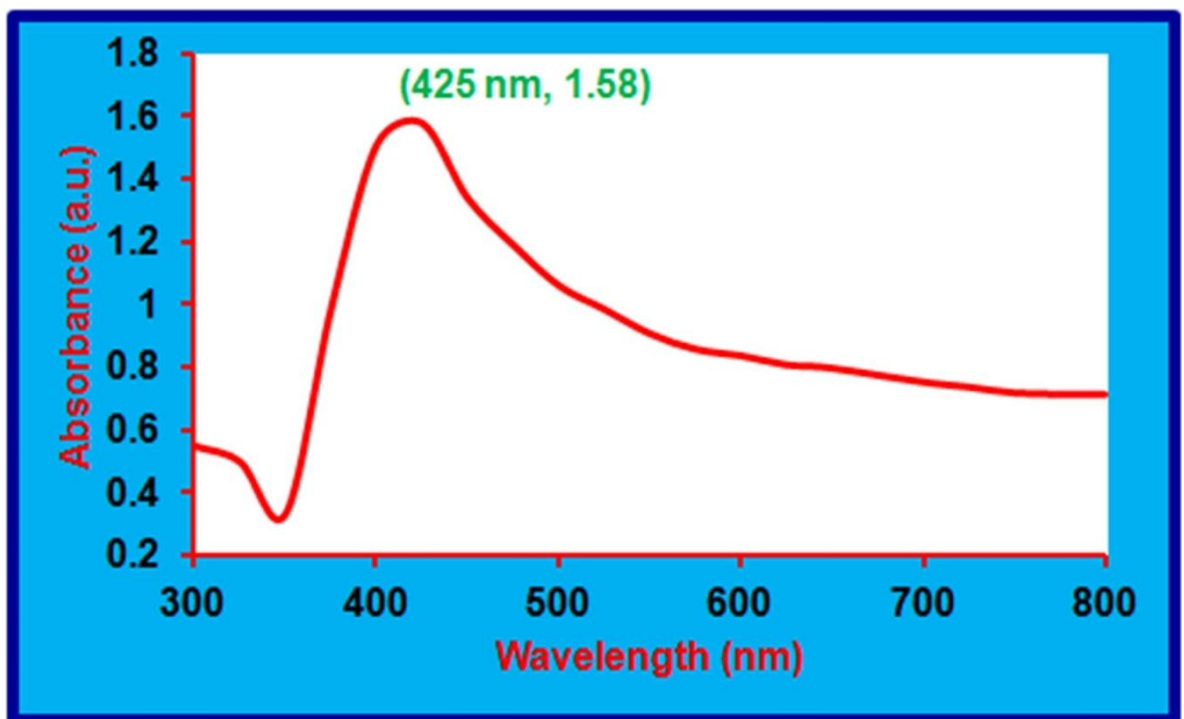

**Fig.4** UV-Vis spectrum of bio-synthesised silver nanoprticles showing SPR peak

When, the particle size is increased, the absorption spectrum becomes narrower with increase in the intensity of the band and decrease in the bandwidth (Barnicket *et al*., 1992; Vivekanandhan *et al*., 2012). There is a linear variation of the half-width with the inverse of the particle diameter (Chandan *et al*., 2006). Petit and co-workers presented such a linear relationship and estimated the size of silver nanoparticles fromthe full width at half-maximum (FWHM) of the absorption spectrum peak using correlation (Mie 1908; Cason *et al*. 2000)

$$FWHM = 50 + 2300/D \quad \ldots\ldots\ldots\ldots\ldots \quad (1)$$

Where FWHM is in nm and D - the particle diameter is in angstrom. FWHM value 100 nm was derived from UV-Vis spectrum and it was applied in eq.1. The particle size of the sample 5 nm was calculated, accordingly (Andrews and Ozin 1986).

Nanoparticles mediated plant transformation has the potential for genetic modification of plants for further improvement. Specifically, application of Nanoparticles technology in plant pathology targets specific agricultural problems in Plant-pathogen interactions and provide new ways for crop protection. There is thus an urgent need for research on interactions between Nanoparticles and environmental matrices (water, sediments and soils) and eco-toxicity studies that take into account the anticipated modifying effect of such matrices on uptake in organisms and toxicity. The interaction between nanoparticles and PGPR bacteria at industrial level may be a step of success towards Green Revolution. Such biologically synthesized, very tiny functional nanoparticles are economically cheap, easy downstream processing and environmentally safe, because they are naturally encapsulated by fungal protein which are water soluble. Such biologically engineered silver nanoparticle synthesized in the cell free filtrate may be directly used in biomedical, engineering, and agricultural sector. The extracellular synthesis of nanoparticles is advantageous from the perspective of large scale production and downstream processing. The obtained results are promising enough to pave the environmentally benign nanoparticle synthesis processes without use of any toxic chemicals and also envision the emerging role of endophytes towards synthesis of nanoparticles.

## 4. Conclusions

In conclusion, we have reported the simple biological way for synthesizing the silver nanoparticles using the culture supernatant of *Pseudomonas sp.* ARS22. The present investigation indicates the extracellular synthesis of highly stable silver nanoparticles. The results of FTIR suggested that the protein might have played an important role in the stabilization of silver nanoparticles. These study results demonstrated that the Plant growth promoting bacteria *Pseudomonas sp.* ARS22 is a cheap and environment-friendly bio-resource for the synthesis of silver nanoparticles with antibacterial activity. Considering the significance of Plant growth promoting bacteria an agriculturally important microbe, their utilization to synthesize AgNPs with potent antibacterial properties can certainly provide an alternate means for plant protection. Further studies are required on fundamental understanding of the mechanism of nanoparticles synthesis at cellular and molecular levels.

Silver nanoparticle has always been an excellent antimicrobial agent. The exclusive physical and chemical properties of silver nanoparticles always intensify the efficacy of silver. Although there are numerous of mechanisms which credited to the antimicrobial activity shown by silver nanoparticles, but still the actual mean is not fully known. Chemical and physical approaches of silver nanoparticle synthesis were being used over the eras, but they are found to be costly and sometime slightly toxic too. The applications of silver nanoparticles are diverse and numerous but the most anticipated characteristic is their antimicrobial and anti-inflammatory efficacy. The remarkably strong antimicrobial activity is the major direction for expansion of nano-silver products. Examples are food packaging materials and food supplements, odour-resistant textiles, electronics, household appliances, cosmetics and medical advices, water disinfectants and room sprays.

### Acknowledgments
Corresponding author is thankful to CSIR for providing RA fellowship and extremely thankful to AIRS, Jawaharlal Nehru University, New Delhi for providing FTIR, SEM, TEM facilities. We are thankful to Amity University, Noida for providing us all required things for accomplishment of this research. Also, thank to staff & management of Kalasalingam University, Krishnankoil, India for their valuable assistances and encouragements during this work.